\documentclass{ws-ijmpe}

\def\w{\omega}

\def\ri{r_{i}}

\begin{document}

\markboth{Dotti, Gleiser \& Ranea-Sandoval}{Instabilities in Kerr Spacetimes}

\catchline{}{}{}{}{}

\title{\titlefont{INSTABILITIES IN KERR SPACETIMES}}

\author{{\authorfont{DOTTI, GUSTAVO and GLEISER, REINALDO J.}}}

\address{ Facultad de Matem\'atica, Astronom\'{\i}a y F\'{\i}sica (FaMAF),
Universidad Nacional de C\'ordoba and  
Instituto de F\'isica Enrique Gaviola, CONICET.\\
Ciudad Universitaria, (5000) C\'ordoba, Argentina.}

\author{\authorfont RANEA-SANDOVAL, IGNACIO F.}
\address{Grupo de Gravitaci\'on, Astrof\'isica y Cosmolog\'ia, Facultad de Ciencias Astron\'omicas y Geof\'{\i}sicas,
 Universidad Nacional de La Plata. Paseo del Bosque S/N (1900), La Plata, Argentina.\\
 CONICET, Rivadavia 1917, 1033, Buenos Aires, Argentina}

\maketitle

\begin{history}
\received{(received date)}
\revised{(revised date)}
\end{history}

\begin{abstract}
We  present a generalization of previous results regarding the stability under gravitational perturbations 
of nakedly singular super extreme Kerr spacetime and Kerr black hole interior beyond the 
Cauchy horizon. To do so we study solutions to the radial and angular Teukolsky's equations with 
different spin weights, particulary $s=\pm 1$ representing electromagnetic perturbations, $s=\pm 1/2$ 
representing a perturbation by a Dirac field and $s=0$ representing perturbations by a scalar field.
 By analizing the properties of radial and angular eigenvalues we prove the existence of an infinite 
family of unstable modes.
\end{abstract}

\section{Introduction}

The most general stationary black hole in electrovacuum is the Kerr-Newman one. In Boyer-Lindquist 
coordinates its line element reads:
\begin{eqnarray} \label{kerr}ds^2 = - \frac{(\Delta-a^2 \sin^2 \theta) }{\Sigma} dt^2 -2a\sin^2
\theta \frac{(r^2 + a^2 - \Delta)}{\Sigma}
dt d\phi \nonumber \\ 
+ \left[ \frac{(r^2+a^2)^2 - \Delta a^2 \sin^2 \theta}{\Sigma} \right] \sin^2 \theta d\phi^2
+
\frac{\Sigma}{\Delta} dr^2 + \Sigma  d\theta^2 ,
\end{eqnarray}
where $\Sigma = r^2 + a^2 \; \cos ^2 \theta$ and $ \; \Delta = r^2-2Mr + a^2 + Q^2$.

The parameters $M$, $a$ and $Q$ correspond to the mass, the angular momentum per unit mass and the electric charge respectively.

Depending on the nature of the roots of the function $\Delta$, the above solution represents a black hole or a naked singularity.

The stability against gravitational perturbations of spacetimes presenting naked singularities were studied: for negative mass  Schwarzschild, \cite{dottigleiser,dottigleiser2} for super charged Reissner-Nordstr\"om \cite{doglepu,dottigleiser3} and for super extreme Kerr. \cite{doglepu,doglersv,doglerspv} The results obtained in these works shed some light into Penrose's Cosmic Censorship Conjeture (CCC) showing systematically the existence of unstable modes. Both the Reissner-Nordstr\"om and the Kerr black holes have a two horizon structure through which spacetime can be extendend, in this process new regions isometric to $I: r > r_o$, $II: r_i < r < r_o$ and $III: r < r_i$ arise ad infinitum.  The inner and outer horizons are located at $r_i   =  M - \sqrt{M^2-a^2 - Q^2}$ and $r_o   =  M + \sqrt{M^2-a^2 - Q^2}$.

 In this work we will focus in Kerr's spacetime because it is the most relevant from an astrophysical point of view.  


The main result of this work is the generalization of the results on gravitational perturbations in 
\cite{doglerspv} to scalar, spinor and Maxwell fields, shown to be unstable in the Kerr naked singularity and 
black hole interior.

\section{Teukolsky Equations \label{te}}

 Teukolsky's equation uses a master variable $\Phi_s$ to describe  scalar, spinor, Maxwell
 and linear gravitational fields, for which $|s|=0,1/2,1$ and $2$ respectively.
 Solutions are separable, 
\begin{equation}
\Phi_s=
R_{\omega,m,s}(r) S^m_{\omega,s}(\theta) \exp(im\phi) \exp(-i\omega t),
\end{equation}
 and the equations reduce to  a coupled system for $S$ and
$R$, \cite{teukolsky} 
\begin{eqnarray} \label{ta}
{1\over \sin\theta} {d \over d\theta}\left(\sin\theta {d S\over
d\theta}\right)+F_\theta(\theta,a,\omega,m,s,E)S &=& 0 \\
\label{tr}
\Delta {d^2 R \over dr^2} +(s+1) {d\Delta\over dr} \;{dR\over dr} +F_r(r,a,\omega,m,s,E)R &=&0 .  
\end{eqnarray}

This formalism was used to establish  the modal stability under gravitational perturbation of 
region I of a Kerr black hole. \cite{stable}
In what follows we will  prove that region III  of an
$a<M$ Kerr spacetime,  as well as
 the nakedly singular $a>M$ Kerr  spacetimes,
 are unstable under gravitational perturbations and all kinds of linear fields, 
as described above for different $s$ values, generalizing our previous work . \cite{doglerspv}


 \subsection{Spin Weighted Spheroidal Harmonics (SWSH's)}
 In this section we gather important aspects of the {\it SWSH}'s eigenvalues, these hold 
for every  $a$ and $M$. 
 In the axial case $m=0$ the $E$ eigenvalue for  large purely imaginary $\w$
  behaves as $(\ell=0,1,2...)$: \cite{bertilong}
 \begin{equation} \label{arrivaberti}
 E_{\ell , s}(a\w)|_{a\w=ik} = (2 \ell  + 1) k + {\cal O} (k^0), \;\;\; \text{ as }  k \to \infty
 \end{equation}
and in the $a \w =0$ limit, independently of the value of $s$ as:
\begin{equation} \label{abajoberti}
E_{\ell , s}(a\w)|_{a\w=0} = (\ell+|s|)(\ell+|s|+1).
\end{equation}

\subsection{Teukolsky Radial Equation}

As was shown for the gravitational case,\cite{doglersv} equation (\ref{tr}) can generally
be cast into a Schr\"odinger-like form, so (\ref{tr}) can be written as:
\begin{equation} \label{scho}
{\cal H} \psi := - \psi'' + V \psi = -E \psi.
\end{equation}
%
with primes denoting derivatives with respect to the radial variable $r^*$ defined in \cite{doglersv}.
The form of the $E$ independent potential $V$ for axial perturbations ($m=0$) 
with purely imaginary frequencies ($\omega = ik/a$), those relevant to unstable modes, is:
\begin{equation} \label{v}
V(r,k):=k^2 V_2(r,k) + k V_1(r,k) + V_0(r,k)
\end{equation}

Understanding the behaviour of the potential for different values of $k$ is important 
 when describing the nature of the eigenvalues of ${\cal H}$. $V_2(r,k)$ is independent 
of the value of the spin weight $s$, so for large values of $k$ the potential behaves like $V_2(r,k)$ 
and so would the eigenvalues of ${\cal H}$. Particulary different is the situation for small values of $k$: 
for $k \ge 0$, $V$   is bounded from below, but its minimum is not a continuous function of $k$ at $k=0$.
%
%
%
%
%
%
This fact is a consequence of the appearence of a second minimum (for $|s|=\pm 1,\pm 2 $) in the function
 $V(r,k \ne 0)$ that moves towards $\infty$ as $k \to 0^{+}$ that is not present in the simpler
 $V(r,k=0):=V_0(r)$ function. So we have that:
 \begin{equation} \label{vm}
 \min \{ V(r,k=0), r \in {\mathbb R} \}  =  1/2 - s^2  , \lim_{k \to 0^+}  \min \{ V(r,k), r \in {\mathbb R} \} =  1/4 -s^2,
 \end{equation}

\section{Results}
\subsection{\label{proof} Unstable Modes of the Kerr Naked Singularity}

 The operator ${\cal H}$ in (\ref{scho}) is self-adjoint in the Hilbert space of square integrable functions of 
$r^*$  with
hermitian product  $\langle \alpha | \beta \rangle := \int \overline{\alpha} \; \beta \; dr^*$. \cite{futuro}
 Given that  $V$ is smooth,  bounded from below, and
 \begin{equation} \label{a1}
 V \sim  \left( \frac{Mk e^{r^*}}{a} \right) ^2 ,  |r^*| \to \infty,
 \end{equation}
the  spectrum of the self-adjoint
  operator ${\cal H}$  is fully discrete and has a lower bound. A careful analysis
   indicates that the square integrable eigenfunctions of ${\cal H}$ behave
as
\begin{equation} \label{ae}
\psi \sim \begin{cases} e^{- \frac{rk}{a}} \left( \frac{M}{r} \right) ^ {\frac{3}{2} + 3s +\frac{2kM}{a}} \left( 1 + {\cal O} (M/r) \right)  & , r \to \infty \\
                    e^{ \frac{rk}{a}} \left( \frac{M}{r} \right) ^ {\frac{3}{2} + s -\frac{2kM}{a}} \left( 1 + {\cal O} (M/r) \right) & , r \to -\infty . \end{cases}
\end{equation}
To obtain information about the fundamental energy of the radial Hamiltonian (\ref{scho}), we need to analyze
its potential (\ref{v}). 
One can show that there is  an interval $r_1(M)< r < r_2(M)<0$ where
  $V_2$ is negative.
Using a smooth test function 
supported in this interval 
we can then show that if $-\epsilon_o(k)$
is the lowest  eigenvalue of ${\cal H}$ then
\begin{equation} \label{radiale}
\epsilon_o(k=0^+)   <  -\frac{1}{4} + s^2
, \hspace{1cm}
 \epsilon_o(k) > \frac{1}{2} | \langle \psi |V_2 |\psi \rangle | k^2 , \; \; k > k_c.
\end{equation}
From (\ref{arrivaberti}), (\ref{abajoberti}) and (\ref{radiale}) it is clear that, for any
$\ell > \ell_o$ ($\ell_o$ a function of $s$), the curves $\epsilon_o(k)$ and  $E_{\ell}(a\w)|_{a\w=ik}$ intersect at some
$k_{\ell} > 0$.

 This implies that there is a common eigenvalue $E =  E_{\ell}(a\w)|_{a\w=ik_{\ell}}
=  \epsilon_o(k_{\ell})$ for the radial and angular Teukolsky equations and
(infinitely many!)
 unstable  perturbations
$ \Phi_{s}(t,r,\theta)=S_{(\ell,m=0,k_o)}(\theta)  \Delta^{-(\frac{2s+1}{4})} \psi_o^{(k_{\ell})}(r) \exp(k_{\ell} t/a)$ with
$\psi_o^{(k_{\ell})}(r)$ behaving as in (\ref{ae}).
\subsection{\label{proof2} Unstable Modes for Region III of a Kerr Black Hole.}
The calculations  above can be adapted to deal with perturbations in the {\em interior} region $r< \ri\}$
of a Kerr black hole.  The extreme and
 sub-extreme cases require separate treatments, we present the results for the extreme case (the subtleties in the subextreme case can be seen in \cite{doglersv}).
Using a similar approach to that used in the superextreme case we get that:
 \begin{equation} \label{av2}
V \sim \begin{cases}  4k^2 \exp(2r^*)  & , r^* \to \infty \\
                k^2 \exp(-2r^*)     & , r^* \to -\infty , \end{cases}
\end{equation}
then the  spectrum of the self-adjoint
  operator ${\cal H}$  is again fully discrete and has a lower bound. The eigenfunctions behave as
 \begin{equation} \label{aee}
\psi \sim \begin{cases} \left(\frac{M}{M-r} \right)^ {2k-s-\frac{1}{2}} \; \exp \left[-2k \left(\frac{M}{M-r} \right) \right]
 \left( 1 + {\cal O} (\frac{M-r}{r}) \right)  & , r \to M^- \\
                   \left( \frac{M}{r}\right)^{\frac{1}{2}-2k-s} \; \exp \left[ \frac{rk}{M} \right]  \left( 1 + {\cal O} (M/r) \right) & , r \to -\infty  \end{cases}
\end{equation}
The argument of instability in the super extreme case goes through without modifications
and thus can be used again to obtain the bound (\ref{radiale}).

\section*{Acknowledgements}

This work was supported in part by grants form CONICET, UNC and UNLP. GD and RJG are supported by CONICET. IFRS is a fellow of CONICET. IFRS wish to acknowledge some useful discussions with Professor H. Vucetich.

\end{document}